\documentclass[letterpaper, 10 pt, conference]{ieeeconf}  

\IEEEoverridecommandlockouts                              
\title{\LARGE \bf
On a hierarchical control strategy \\
for multi-agent formation without reflection
}

\author{Toshiharu Sugie,  Brian D. O. Anderson,  Zhiyong Sun and Huichao Dong 
\thanks{T. Sugie and H. Dong are with Department of Systems Science, Kyoto University, Sakyo-ku, Kyoto 606-8501, Japan. Emails:
        {\tt\small  sugie@i.kyoto-u.ac.jp, dong.hucihao@ctrl.sys.i.kyoto-u.ac.jp}
         }%
\thanks{B.~D.~O. Anderson is with Hangzhou Dianzi University, Hangzhou, China, The Australian National University and Data-61 CSIRO,  Canberra, ACT 2601, Australia. Email:
  {\tt\small Brian.Anderson@anu.edu.au}. 
  }%
\thanks{Z. Sun was with Research School of Engineering, The Australian National University,  Canberra, ACT 2601, Australia. He is now with Department of Automatic Control, Lund University, Sweden.  Email:
  {\tt\small sun.zhiyong.cn@gmail.com}.
  }
}
\usepackage{graphicx}
\usepackage{epstopdf}
\usepackage{amsmath}
\usepackage{amsfonts}
\usepackage{amssymb}
\usepackage{mathrsfs}
\usepackage{color}
\usepackage{subcaption}

\newtheorem{remark}{\textbf{Remark}}

\begin{document}

\maketitle
\thispagestyle{empty}
\pagestyle{empty}

\begin{abstract}
This paper considers a formation shape control problem for point agents in a two-dimensional ambient space, where the control is distributed, is based on achieving desired distances between nominated agent pairs, and avoids the possibility of reflection ambiguities. 
This has potential applications for large-scale multi-agent systems having simple information exchange structure. One solution to this type of problem, applicable to formations with just three or four agents, was recently given by considering a potential function which consists of both   distance error and   signed triangle area terms. However, it seems to be challenging to apply it to formations with more than four agents. This paper shows a hierarchical control strategy which can be applicable to any number of agents based on the above type of potential 
function and a formation shaping incorporating a grouping of equilateral triangles, so that all controlled distances are in fact the same. A key analytical result and some numerical results are shown to demonstrate the effectiveness of the proposed method.
\end{abstract}

\section{INTRODUCTION}

Formation shape control for multi-agent systems is one of the most actively studied topics due to its potential in various applications and theoretical depth. Surveys of formation shape control are found in \cite{AYFH2008} and \cite{OPA2015}.  According to the sensing capability and agent-interaction topology, most of the existing methods can be classified as (a) Position-based control, (b) Displacement-based control, and (c) Distance-based control (see \cite{OPA2015}). In the case of (a), we need the absolute position of each agent, often to a high accuracy, so sensors (like GPS) could be expensive. In the case of (b), most of the existing works require that all the different local coordinate systems associated with each agent should be aligned with a global coordinate system. This might be difficult from the implementation viewpoint. In contrast, in the case of (c), each agent only requires the relative position information in its own local coordinate system. Hence, at least in part because of the implied saving in sensor requirements relative to (a) and (b), distance-based formation control has attracted a considerable attention recently
 (see references in  \cite{ASSAS2017}, \cite{SAS2015}, \cite{SAS2018}, \cite{OPA2015}, \cite{KBF2009}, \cite{Sun2016}). One major drawback is that there can be many undesirable equilibria when using the gradient control laws that are typically suggested. Because of the overall  system's nonlinear control nature, it is not trivial to guarantee the convergence to the desired formation shape from all or almost all initial conditions. Also, the approach may require more inter-agent interactions to achieve a   desired rigid formation compared to the case of (b). For example, if the target system is described by four agents with five edges (i.e., inter-agent interactions) and their lengths in a two dimensional space, the system is \textit{not} uniquely specified up to congruence, due to the possibility of reflection ambiguities, either of
individual triangles in the formation or the whole formation.  Six edges are necessary, if individual triangle reflection ambiguities are to be avoided, while a reflection ambiguity remains for the whole
formation.

\begin{figure}[htb]
 \begin{center}
  \includegraphics[width=0.3\linewidth]{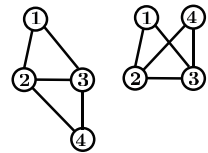} \end{center}
 \caption{An example of flip ambiguities}
 \label{fig:zero}
\end{figure}

Recently, Anderson et al. (\cite{ASSAS2017}) introduced a new approach for formation shape control to address this issue, initially for a triangular formation or a formation with four agents,   which utilizes a potential function including not only the distance errors but also a signed area term. To the best of our knowledge, the control scheme proposed in \cite{ASSAS2017} is the only solution to resolve reflection ambiguity in distance-based formation systems both for individual triangles and for the whole formation.  The resultant controller requires only the relative position measurements in each 
 agents'  local coordinate frame. Nevertheless, it is able to prevent the occurrence of flip or reflection ambiguity. More precisely, in the three-agent case, specifying a triangle in terms of the lengths of its sides means that two formations, one being the reflection of the other, meet the distance criteria. The proposed algorithm in \cite{ASSAS2017} however enables a particular one of these possibilities always to be secured. Specifying a four-agent formation by the lengths of the sides of two triangles gives rise to further flip ambiguities (see Fig. \ref{fig:zero}) associated with individual triangles, and again a unique formation among the several possibilities can always be obtained.
 
%
%
This idea could be a powerful tool of formation control for large-scale multi-agent systems. Unfortunately, the paper \cite{ASSAS2017} discussed the formation control for three-agent and four-agent cases only, and it seems to be challenging to handle the general case despite the potential which the new approach inherently has.

The purpose of this paper is to provide a formation control strategy which is applicable to any number of agents based on the potential function idea of   \cite{ASSAS2017}.  As the first step, this paper discusses the case of formations obtained by combining equilateral triangles, with the requirement that flip ambiguities be avoided. A key analytical result is shown, and some numerical results are given to demonstrate the effectiveness of the proposed method. In the paper we do not consider collision avoidance between individual agents, which will be a future research topic (while available techniques in e.g., \cite{olfati2003,zhao2018} can be applied to the current control framework).

The layout of the paper is as follows. In Section II, we provide a formal problem statement, and in Section III describe the proposed method, a key step being the demonstration of a hierarchical construction of control laws. Sections IV and V include further simulation results and concluding remarks, respectively.

\section{PROBLEM SETTING}
Consider a multi-agent formation system consisting of $n$ point agents in a 2-dimensional space which are governed by
the equations
\begin{eqnarray}
 \dot{p}_i (t)=  u_i (t) \hspace{5mm}  i \in {\cal V},
\end{eqnarray}  
where  $p_i(t) \in {\bf R}^2$ and $u_i(t) \in {\bf R}^2$ are the state and the input of agent $i$, and ${\cal V}:=\{1,2,\cdots,n \}$ denotes the set of all agents. The information structure exchanged between agents is described by an undirected graph 
${\cal G}=({\cal V},{\cal E})$, where ${\cal E} \subset{ {\cal V} \times {\cal V}}$ denotes the set of edges. For example, $(i,j)\in {\cal E}$ implies that two agents $i$ and $j$ exchange information with each other.
Also, ${\cal {N}}_{i} \subset  \cal V$ denotes the set of all neighbors of agent $i$ which is defined by 
\begin{eqnarray*}
{\cal{N}}_{i}  := \{ j\in{\cal V}:(i,j)\in {\cal E}, i\neq j \}. 
\end{eqnarray*} 
Agent $i$ detects the relative position of the neighbor agent $j$ ($j \in {\cal{N}}_{i}$) in its local coordinate frame, and the control input $u_i(t)$ should be of the form of
\begin{eqnarray*}
u_i(t) := f_{i}((p_i(t)-p_j(t)) _{j\in{\cal{N}}_{i} }). 
\end{eqnarray*} 
Define the collective states of all agents by
\begin{eqnarray*}
p(t) := [p_1^T(t), p_2^T(t), \cdots, p_n^T(t)]^T,
\end{eqnarray*}
and let ${\cal P}$ denote the set of all $p \in {\bf R}^{2n}$ which achieves the \textit{desired formation}, which is specified \textit{a priori} up to translation and rotation.  The control objective is to find $u_i(t)~(i=1 \sim n)$ which aims to drive
\begin{eqnarray*}
\lim_{t \to \infty} p(t)  = p(\infty) \in {\cal P}.
\end{eqnarray*}

\begin{figure}[htb]
 \begin{center}
  \includegraphics[width=0.3\linewidth]{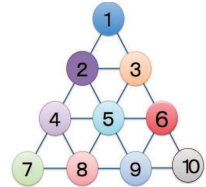} \end{center}
 \caption{An example of graph structure of a multi-agent system}
 \label{fig:one}
\end{figure}

As the first step, we assume that the graph is a triangulated Laman graph \cite{CBB2017} and the desired formation consists of connected equilateral triangles only. Such triangulated graphs can be constructed by the operation of vertex extensions in a Henneberg sequence (see the survey \cite{AYFH2008}). Note that while rotation and translation for the formation are acceptable, the position of two agents is not exchangeable. In particular,  the flipping (or reflection) of each triangle is not acceptable. More precisely, the set ${\cal P}$ will be described 
as set out in detail below:

If three agents $\{i,j,k\}$ satisfy
\[
\{(i,j), (j,k), (k,i) \} \in {\cal E}, 
\]
they are said to form a clique. The set of all such triples $(i,j,k)$ is denoted by ${\cal C}$. For example, 
${\cal C}$ is given by
\begin{eqnarray*}
{\cal C} = &\{ (1,2,3),~(2,3,5),~(2,4,5),~((3,5,6)\\
&(4,7,8),~(4,5,8),~(5,8,9),~(5,6,9),~(6,9,10) \}
\end{eqnarray*}
in the case of the graph shown in Fig. \ref{fig:one}. For each clique ($i,j,k$), we define the signed area $Z_{i,j,k}$ by
\begin{eqnarray}
Z_{i,j,k} & := & \frac{1}{2}\mbox{det}\begin{bmatrix}
     1&1&1 \\
       p_i&p_j&p_k\\
    \end{bmatrix}.
  \end{eqnarray}
 It is easy to see that $|Z_{i,j,k}|$ equals the area of the triangle ($i,j,k$), and $Z_{i,j,k}$ is positive if the three agents' positions, $p_i$, $p_j$ and $p_k$, are located in a counterclockwise ordering; otherwise it is negative. Then, ${\cal P}$ consists of $p \in {\bf R}^{2n}$ satisfying the following two conditions:
 \begin{itemize}
\item[(A)]
 $||p_i-p_j||=d^*$, ~~ $ \forall  (i,j) \in {\cal E}$ 
\item[(B)]
$Z_{i,j,k}=Z_{i,j,k}^{*}$,~~ $ \forall  (i,j,k) \in {\cal C}$ 
 \end{itemize}
where $d^*$ is the given desired distance between two agents and $Z_{i,j,k}^{*}$ denotes the given desired \textit{signed} area.
Obviously these specifications describe a formation which is unique up to translation and rotation, but preclude any flipping ambiguity. 
\begin{remark}
Note that Olfati-Saber \cite{O-S2006} discussed a multi-agent flocking system with $\alpha$-lattice structure. 
The target formation in \cite{O-S2006} looks very similar to the triangulated formation discussed in this paper.  However, the method in \cite{O-S2006} appears restricted \textit{a priori} to working with formations comprising assemblages of  triangles and cannot control the flipping for the specified agents. Our choice of equilateral triangles is principally for illustrative purposes (although more precise numerical results can
also be obtained). Furthermore, the paper \cite{O-S2006} considered a different potential function, and discussed double-integrator flocking systems with some general (and local) convergence results, while no reflection issue was considered. In this paper we focus on distance-based formation stabilization without reflection, which is a different problem as compared to that in \cite{O-S2006}. 
\end{remark}


\section{PROPOSED METHOD}
This section gives a solution to the general problem posed prior to the example above.  First, a special case for two/three-agent systems will be analyzed. Then, a hierarchical control strategy will be proposed. 
The three-agent formation case  differs from the treatment in [2], in that two agents are pinned a priori. 

In the following, gradient descent laws are used. 
Since the potential function under discussion is real analytic and the formation system is a gradient-descent system,  individual agents' positions always converge to a critical point of the associated cost function (see \cite{AK2006}).

\subsection{Analysis for the two and three agent cases}

The following result, preliminary to and used in multi-agent problems below, is obtained for a two-agent case. 

{\bf Theorem 1}~ Suppose the formation system consists of two agents $i$ and $j$. If $p_i$ is fixed and $p_j$ is governed by
\begin{eqnarray}
\dot{p_j} =  -\frac{\partial V_{(i,j)} }{\partial{p_j}}, 
\end{eqnarray}
with 
\begin{eqnarray}
V_{(i,j)}  := \frac{1}{4}((||p_i-p_j||^2-(d^{*})^2)^2,
\end{eqnarray}
then $p_j$ converges and all stable equilibria of $p_j$ satisfy 
\begin{eqnarray}
||p_i-p_j||= d^*,
\label{eq:desired-f}
\end{eqnarray}
i.e., $p_j$ converges to a point with the desired distance from $p_i$.

{\bf Proof:}  
Assume 
\begin{eqnarray}
p_i = 
\begin{bmatrix}
     0 \\
      0
\end{bmatrix},~
  p_j=  \begin{bmatrix}
     x\\
     0
    \end{bmatrix},
\label{eq:pjxy}
\end{eqnarray}
without loss of generality. Then, we have
\begin{align*}
 \dot{p_j} & = -(||p_i-p_j||^2-(d^*)^2)(p_j-p_i) \\
&= -(x^2-(d^*)^2)
 \begin{bmatrix}
     x\\
      0
    \end{bmatrix}.
 \end{align*}
 It is obvious that $p_j$ moves along the $x$-axis (i.e., $y =0$ always holds). Therefore, it is enough to discuss
 the behaviour of $p_j$ in the $x$-axis only. 
 It is easy to see that the possible equilibria are $x=0$ and  $x= \pm d^{*}$.

Now we will analyze the stability of each equilibrium point. 
The potential function $V$ (corresponding to the restriction of behaviour to the $x$-axis) is given by
\begin{eqnarray*}
 V= \frac{1}{4} (x^2 - (d^*)^2)^2.
 \end{eqnarray*}
Hence its Hessian $H$ is calculated as
\begin{eqnarray*}
 H= 3x^2-(d^*)^2.
 \end{eqnarray*}
At the origin $x=0$, the above Hessian will be
$H= -(d^*)^2 < 0$. This implies that 
$x=0$ is not stable, while the other two equilibria are (locally) exponentially stable. In fact, no.  No trajectories (except the one that starts from the origin) will converge to the origin. Only the trajectory obtained with an initial condition at the origin will have the origin as its equilibrium point. 
On the other hand, when $x= \pm d^*$ holds, the Hessian becomes $H=2 (d^*)^2>0$.
Hence the equilibria $x= \pm d^*$ are exponentially stable, which proves the theorem.  {\bf (QED)}

Now we consider a three-agent case. We exploit the following potential function proposed by Anderson et al.  \cite{ASSAS2017}.
\begin{eqnarray} \label{eq:potential_Vijk}
\begin{split}
V_{(i,j,k)} =  \frac{1}{4}\Bigl(&( \| p_i-p_j \|^2-{d_{ij}^*}^2)^2\\
&+(\|p_j-p_k \|^2-{d_{jk}^*}^2)^2\\
&+( \| p_k-p_i \|^2- {d_{ki}^*}^2)^2\Bigr)\\
&+ \frac{1}{2}K(Z_{i,j,k}- Z_{i,j,k}^{*})^2,
\end{split}
\label{eq:V-ijk}
\end{eqnarray}
where $d_{ij}^*$ denotes the desired distance between agents $i$ and $j$, 
$d_{jk}^*$ and $d_{ki}^*$ are similarly defined, and $K$ is a positive control gain associated with the signed area term. It was shown in \cite{ASSAS2017} 
that the signed area error term (i.e., $\frac{1}{2}K(Z_{i,j,k}- Z_{i,j,k}^{*})^2$) plays the central role in preventing convergence to a formation which is a flipped version of the desired formation, i.e., one with the same edge lengths but oppositely signed area. Now the following result is obtained, which is one of the main contributions of this paper.

{\bf Theorem 2:}~ Suppose the formation  system consists of three agents $i$, $j$ and $k$.  In addition, agents $i$ and $j$ are fixed, and their pinned positions satisfy $\|p_i-p_j \|=d^*$. Also, assume that agent $k$ is governed by 
\begin{eqnarray} \label{eq:gradient_system_3agent}
 \dot{p_k} = -\frac{\partial V_{(i,j,k)}}{\partial {p_k}},
\label{eq:dot-pk}
\end{eqnarray}
with $d_{ij}^*=d_{jk}^*=d_{ki}^*=d^*$, and the potential function $V_{(i,j,k)}$ is defined in  \eqref{eq:potential_Vijk}. 
Then, (i) if $K > 3/2$ holds, $p_k$ converges globally to the unique correct equilibrium  in ${\cal P}$; (ii) if $-2+2\sqrt{3} < K \leq 3/2$ holds, there exist a stable correct equilibrium in ${\cal P}$ and two unstable incorrect equilibria, and almost all trajectories of $p_k$ converge to the correct equilibrium formation in ${\cal P}$;  
(iii) if $0 < K < -2+2\sqrt{3}$, there exist a locally stable correct equilibrium in ${\cal P}$ and one locally stable incorrect equilibrium not in ${\cal P}$ (as well as unstable saddle equilibrium points). In this case, $p_k$ may converge to an incorrect formation point not in ${\cal P}$ depending on the initial position. 

{\bf Proof:}~
Without loss of generality, we assume $Z_{i,j,k}^{*}\textgreater 0$ and 
\begin{eqnarray}
p_i = \begin{bmatrix}
     -a \\
      0
    \end{bmatrix},
    p_j= \begin{bmatrix}
     a \\
      0
    \end{bmatrix},
    p_k=  \begin{bmatrix}
     x\\
      y
    \end{bmatrix},
\label{eq:pkxy}
\end{eqnarray}
where $a:= d^*/2>0$. From  (\ref{eq:V-ijk}) and (\ref{eq:dot-pk}), we have
\begin{eqnarray*}
\begin{split}
 \dot{p_k} = &-(\| p_i-p_k \|^2-4a^2)(p_k-p_i)\\
& -(\|p_j-p_k \|^2- 4a^2)(p_k-p_j)\\
 &-\frac{1}{2}K(Z_{i,j,k}- Z_{i,j,k}^{*})
\begin{bmatrix}
 0&1\\
 -1&0\\
 \end{bmatrix}
 (p_i -p_j ).
 \end{split}
 \end{eqnarray*}
By substituting (\ref{eq:pkxy})  into the above with 
\begin{eqnarray}
 Z_{i,j,k}=ay, ~~ Z_{i,j,k}^{*}=\sqrt{3}a^2,
\end{eqnarray}
we have
\begin{eqnarray*}
\begin{split}
 \dot{p_k} = &-(x+a)^2+y^2-4a^2)
\begin{bmatrix}
     x+a\\
      y\\
    \end{bmatrix}\\
 &-((x-a)^2+y^2-4a^2)
 \begin{bmatrix}
     x-a\\
      y
    \end{bmatrix}\\
   &-\frac{1}{2}K(ay-\sqrt{3}a^2)
\begin{bmatrix}
     0\\
     2a\\
    \end{bmatrix},
    \end{split}
 \end{eqnarray*}
namely,   
\begin{eqnarray}
 \dot{p_k}  =  
 \begin{bmatrix}
     -2x(x^2+y^2-a^2)\\
      -2y(x^2+y^2-3a^2)+Ka^2(\sqrt{3}a-y)
    \end{bmatrix}.
\label{eq:pkdot} 
 \end{eqnarray}
Based on the above equation, we will compute the equilibria.

First we consider the case of $x=0$. The corresponding second entry 
of $\dot{p}_k$ should satisfy 
\begin{eqnarray*}
      -2y(y^2-3a^2)+Ka^2(\sqrt{3}a-y)=0,
\end{eqnarray*}
which implies 
\begin{eqnarray*}
      (\sqrt{3}a-y)(Ka^2+2y(\sqrt{3}a+y))=0.
\end{eqnarray*}
Hence one equilibrium is given by
\begin{eqnarray}
p_a^*= 
\begin{bmatrix} 
0 \\
\sqrt{3}a
\end{bmatrix}.
\end{eqnarray}
On the other hand, one can observe  that a second equilibrium would exist given a real $y$ satisfying the equation
\begin{eqnarray*}
    Ka^2+2y(\sqrt{3}a+y)=0
 \end{eqnarray*}
However, this equation is equivalent to
\begin{eqnarray}
   (y+\frac{\sqrt{3}}{2}a)^2+(\frac{K}{2}-\frac{3}{4})a^2=0.
    \label{eq:yK}  
 \end{eqnarray}
Hence, if $K > 3/2$,  no other equilibrium point exists. 

If now $x=0$ and $K\leq 3/2$, then the following two equilibria exist.
\begin{eqnarray*}
p_b^*= 
 \begin{bmatrix}
0\\
   \left( -\sqrt{\frac{3}{4}-\frac{K}{2}}-\frac{\sqrt{3}}{2}\right) a\\
    \end{bmatrix},~
  p_c^*=
   \begin{bmatrix}
     0\\
   \left( \sqrt{\frac{3}{4}-\frac{K}{2}}-\frac{\sqrt{3}}{2} \right) a\\
    \end{bmatrix}.
 \end{eqnarray*}
We note that if $K =3/2$ then the above two equilibria reduce to a single equilibrium $p_b^* = [0, -\frac{\sqrt{3}}{2}a]^T$. Now we analyze the stability of each point. From (\ref{eq:V-ijk}) and (\ref{eq:pkxy}), we have
 \begin{align*}
 V_{(i,j,k)}= & \frac{1}{2}x^4+\frac{1}{2}y^4+\frac{9}{2}a^4+x^2y^2-x^2a^2 \\
& -3a^2y^2+\frac{K}{2}a^2y^2+\frac{3}{2}Ka^4-\sqrt{3}Ka^3  y.
\end{align*}
Its Hessian $H$ is given by
\begin{align} \label{eq:Hessian}
 H=\begin{bmatrix}
 6x^2+2y^2-2a^2&4xy\\
 4xy&6y^2+2x^2-6a^2+Ka^2
 \end{bmatrix}.
\end{align}  
When $p_k =p_a^*$ (i.e, $(x,y)=(0,\sqrt{3}a)$) holds, $H$ becomes  
\begin{eqnarray*}
 H=\begin{bmatrix}
 4a^2&0\\
 0&12a^2+Ka^2\\
 \end{bmatrix}.
\end{eqnarray*}  
Therefore, $H$ is positive definite, and thus $p_a^*$ is a stable equilibrium point.

When $p_k=p_b^*$, $H$ is calculated as 
\begin{eqnarray*}
 H= 2a^2 
\begin{bmatrix}
h(K) & 0 \\
0     &  3 h(K) +\frac{1}{2}K
\end{bmatrix},
\\
h(K)  :=  \frac{1}{2}-\frac{K}{2}+\sqrt{\frac{9}{4}- \frac{3K}{2}}.
\end{eqnarray*}
Under the condition $0 \textless K   \leq \frac{3}{2}$, $h(K)$ is a strictly decreasing function with respect to $K$, and it is easy to verify that
\begin{eqnarray*}
h(0)=2, ~ h(-2+ 2\sqrt{3})=0,~ h(\frac{3}{2})=-\frac{1}{4}, 
\end{eqnarray*}
hold. Therefore, when $0<K<-2+ 2\sqrt{3}$ holds, $h(K)$ is positive, and therefore, $H$ is positive definite. This implies $p_b^*$ is a stable equilibrium in this case. When $K = -2+ 2\sqrt{3}$, the Hessian at $p_b^*$ is degenerate (it has a zero eigenvalue and a positive eigenvalue), and the stability of $p_b^*$ is undetermined.   If however 
$-2+ 2\sqrt{3} < K \leq 3/2$ holds, then $p_b^*$ is not a stable equilibrium anymore since the Hessian $H$ at $p_b^*$ has (at least one) negative eigenvalue(s). 

When $p_k=p_c^*$, $H$ is calculated as 
\begin{eqnarray*}
 H= 2a^2 
 \begin{bmatrix}
h_c(K) &0\\
0&3 h_c(K) +\frac{1}{2}K
\end{bmatrix},
\\
h_c(K)  := \frac{1}{2}-\frac{K}{2}-\sqrt{\frac{9}{4}- \frac{3K}{2}}.
\end{eqnarray*}
It is straightforward to show that $h_c(K) < 0$ 
holds if $0<K  \leq 3/2$. Hence $p_c^*$ is an unstable equilibrium.

Second, by returning to \eqref{eq:pkdot}  we consider the case of $x^2+y^2=a^2$ with $x \neq 0$, which makes the first entry of $\dot{p}_k$ in (\ref{eq:pkdot}) zero. Then,  from $\dot{p}_k=0$, we have 
\begin{eqnarray*}
      4a^2y+Ka^2(\sqrt{3}a-y)=0,
 \end{eqnarray*}
from the second entry of (\ref{eq:pkdot}). This implies 
\begin{eqnarray}
    y= \frac{\sqrt{3}Ka}{K-4}.
 \end{eqnarray}
The above and $|y|  < {a}$ yield 
 \begin{eqnarray*}
-1  < \frac{\sqrt{3}K}{K-4}  <  1.
 \end{eqnarray*}
When $K \ge 4$, the above inequalities never hold. In the case of $K< 4$, the above relation 
with $K>0$ is equivalent to 
\begin{eqnarray} \label{eq:condition}
0 <  K  < 2(\sqrt{3}-1).
\end{eqnarray}
The above condition on $K$ ensures the existence of such equilibria. In other words, if $K \geq 2\sqrt{3}-2$, there exists no equilibrium that satisfies the condition $x^2+y^2=a^2$. 

Now we analyze the property of Hessian matrix $H$ at an equilibrium point $p_d^* = [x^*, y^*]^T$ that satisfies ${x^*}^2+{y^*}^2=a^2$. From the general Hessian formula in \eqref{eq:Hessian}, one can obtain the following specific formula for the Hessian matrix at an equilibrium point $p_d^*$ 
\begin{align}
 H_{p_d^*}=\begin{bmatrix}
 4{x^*}^2&4x^*y^*\\
 4x^*y^*&4{y^*}^2 - 4a^2+Ka^2
 \end{bmatrix}.
\end{align}
To ensure that such an equilibrium point $p_d^*$ is stable, there must hold $\text{det} (H_{p_d^*}) >0$ (by observing that $4{x^*}^2>0$ since $x^* \neq 0$). The condition for a stable $p_d^*$ is equivalently stated as
$$4{x^*}^2(4{y^*}^2 - 4a^2+Ka^2) - (4x^*y^*)^2 = 4{x^*}^2a^2(K-4) >0$$
Again, by observing that ${x^*}^2a^2>0$, one must have $K>4$ so that the equilibrium $p_d^*$ is stable (if it exists). However, such a condition $K>4$ contradicts with the condition $0 <  K  < 2(\sqrt{3}-1)$ as derived in \eqref{eq:condition} that ensures the existence of such an equilibrium $p_d^*$. In summary, there exists no stable equilibrium $p_d^*$ satisfying ${x^*}^2+{y^*}^2=a^2$. 

We now summarize the main results for all cases as follows.
\begin{itemize}
\item When $K > 3/2$, there is only one equilibrium $p_a^*$ which is the globally stable, correct equilibrium in ${\cal P}$;
\item When $-2+2\sqrt{3} < K \leq 3/2$, there exist three equilibria, i.e., the correct and almost globally stable equilibrium $p_a^* \in {\cal P}$, and two unstable equilibria $p_b^*, p_c^* \notin {\cal P}$;
\item When $0 < K < -2+2\sqrt{3}$, there exist one  locally stable correct equilibrium $p_a^* \in {\cal P}$ and one locally stable incorrect equilibrium $p_b^* \notin {\cal P}$ (and some saddle points). 
\end{itemize}


The proof is thus complete. 
{\bf (QED)}
\begin{remark}
Because the stable equilibrium points are all associated with
positive definite Hessians (as shown in the proof), convergence actually occurs exponentially fast. Furthermore,  besides the standard gradient-based formation system in \eqref{eq:gradient_system_3agent}, one can also include a general control gain $\kappa$ for the gradient control law in the form of  $\dot{p_k} = -\kappa \frac{\partial V_{(i,j,k)}}{\partial {p_k}}$ to adjust the convergence rate (but with no effect on the equilibria). 
\end{remark}

Note that the paper [2] showed that three agents achieve a correct formation with correct distances and signed area for arbitrary $d_{ij}^*$  for a large enough  $K$ when all agents can move freely. However, we can show that this is not the case for a general triangular formation (as opposed to an equilateral triangle) if two agents are pinned. In other words, the existence of pinned agents makes a big difference for the convergence analysis. In this paper we focus on the special case of equilateral triangles, for which a large enough $K$ can ensure a correct convergence to a desired shape. Theorem 2 shows that, even if two agents are pinned, the correct formation can be achieved by choosing large enough $K$ when $d_{ij}^*=d_{jk}^*=d_{ki}^*$ holds.

\begin{figure}[htb]
 \begin{center}
  \includegraphics[width=0.55\linewidth]{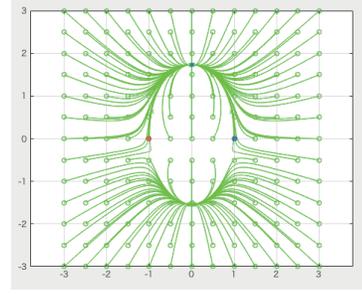} \end{center}
 \caption{Trajectories of agent $k$ in case of $K=0.6$}
 \label{fig:two}
\end{figure}

\begin{figure}[htb]
 \begin{center}
  \includegraphics[width=0.55\linewidth]{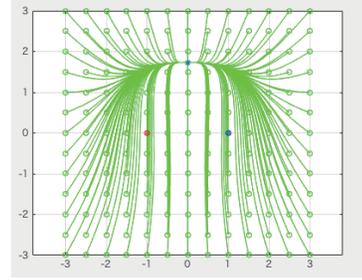} \end{center}
 \caption{Trajectories of agent $k$ in case of $K=20$}
 \label{fig:three}
\end{figure}

It is straightforward to illustrate the behaviour of agent $k$ governed by (\ref{eq:dot-pk}). 
Let agents $i$ and $j$ be located at $p_i=(-1,0)^T$ and $p_j=(1,0)^T$, respectively.   
The target position of agent $k$ is set to be $(0, \sqrt{3})^T$.  
Namely, $d^*=2$ and $Z_{i,j,k}^*=\sqrt{3}>0$. Agents $i$, $j$, $k$ are expected to form 
an equilateral triangle with side length $2$, and $i$, $j$ $k$ should be ordered in a  counter-clockwise 
direction.  
Fig. \ref{fig:two} shows the trajectories of $p_k(t)$ starting from various points marked at small circles,
when $K=0.6$. 
In this case, some trajectories converge to the correct point  $(0, \sqrt{3})^T$, but others do not. On the other hand, 
Fig \ref{fig:three} shows the trajectories of $p_k(t)$ when $K=20$. All trajectories converge to the correct target point.    
These results are  consistent with Theorem 2. 

\subsection{Hierarchical control strategy for the $n$-agent case}

We define the control input as
\begin{eqnarray}
 u_i = -\frac{\partial V_{i}}{\partial {p_i}},
\label{eq:u-i}
\end{eqnarray}
where $V_i$ is chosen as shown below. 

We show how to choose $V_i$  for the system shown in Fig. \ref{fig:one} as an example.
In this case, there exist 9 equilateral triangles, and they should satisfy
\begin{eqnarray*}
Z_{1,2,3}^{*}=Z_{3,2,5}^{*}=Z_{5,2,4}^{*}=Z_{3,5,6}^{*}  \\
=Z_{5,4,8}^{*}=Z_{6,5,9,}^{*}=Z_{8,4,7}^{*}=Z_{6,9,10}^{*} >0.
\end{eqnarray*}

Consistently with Fig. \ref{fig:one}, we choose $V_i$ for each agent as follows:

Layer 1: ~agent 1   (which is stationary)
\[
V_1 \equiv 0.
\]

Layer 2: ~agent 2 (which is to be a fixed distance from agent 1)
\[
V_2 = V_{(1,2)}.
\]

Layer 3: ~agent 3 (which is to form an equilateral triangle with agents 1 and 2)
\[
V_3= V_{(1,2,3)}.
\]

Layer 4: ~ agents $\{4,5,6 \}$ (which are to form equilateral triangles with agents 2 and 3, then 2 and 4, then 3 and 5)
\[
V_5=V_{(3,2,5)}, ~V_4=V_{(5,2,4)},~V_6=V_{(3,5,6)}.
\]

Layer 5: ~agents  $ \{7,8,9,10\}$
\[
V_8=V_{(5,4,8)},~V_9=V_{(6,5,9)},~ V_7=V_{(8,4,7)}, ~V_{10}=V_{(6,9,10)}.
\]

Note that the upper layer agents are never affected by any lower layer agents. Agent 1 stays stationary throughout the whole process, and agent 2 positions itself at the correct distance from agent 1 (direction being irrelevant). Once these two agents are fixed, agent 3 moves to the unique correct point. Then agent 5 approaches to the point which forms the correct triangle $(3,2,5)$. Repeating the behaviour of $p_k$ in Theorem 2, all agents achieve the target formation. The number of agents can be arbitrarily large.

By invoking the stability theory for cascaded systems (see Corollary 9.3 in \cite{Terrell2009}), one can show that if the trajectory of each agent remains bounded, with a large enough gain $K$ from Theorem 2
all trajectories will converge to the correct formation. A rigorous proof for this fact will be reported in an extended version of the paper. 

\section{Simulation}

\begin{figure}[t]  
  \centering
  \begin{subfigure}{.48\columnwidth}
    \centering
    \includegraphics[width=\linewidth]{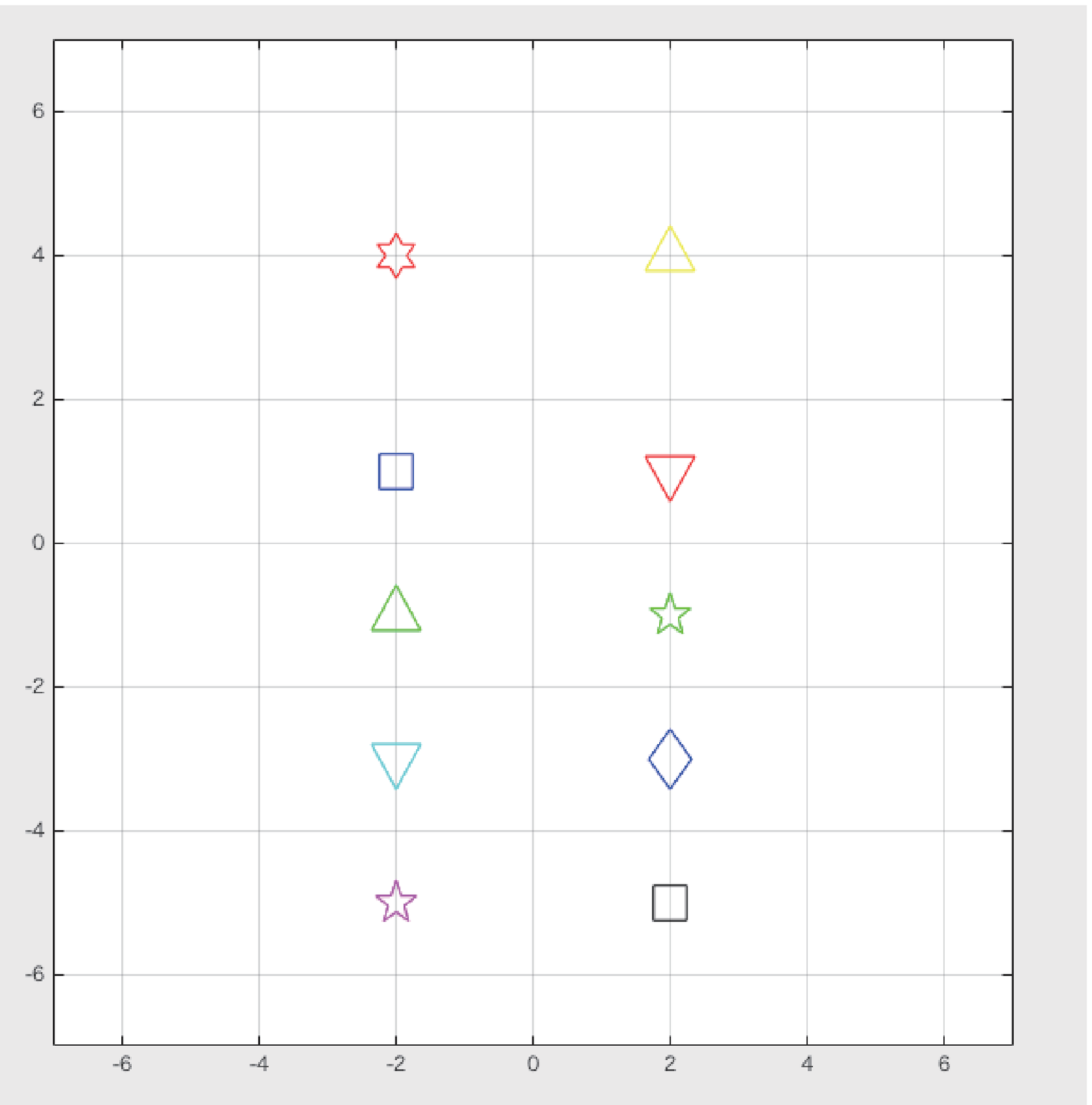}  
    \caption{$t=0$[$s$]}
  \end{subfigure}%
  \hfill
  \begin{subfigure}{.49\columnwidth}
    \centering
    \includegraphics[width=\linewidth]{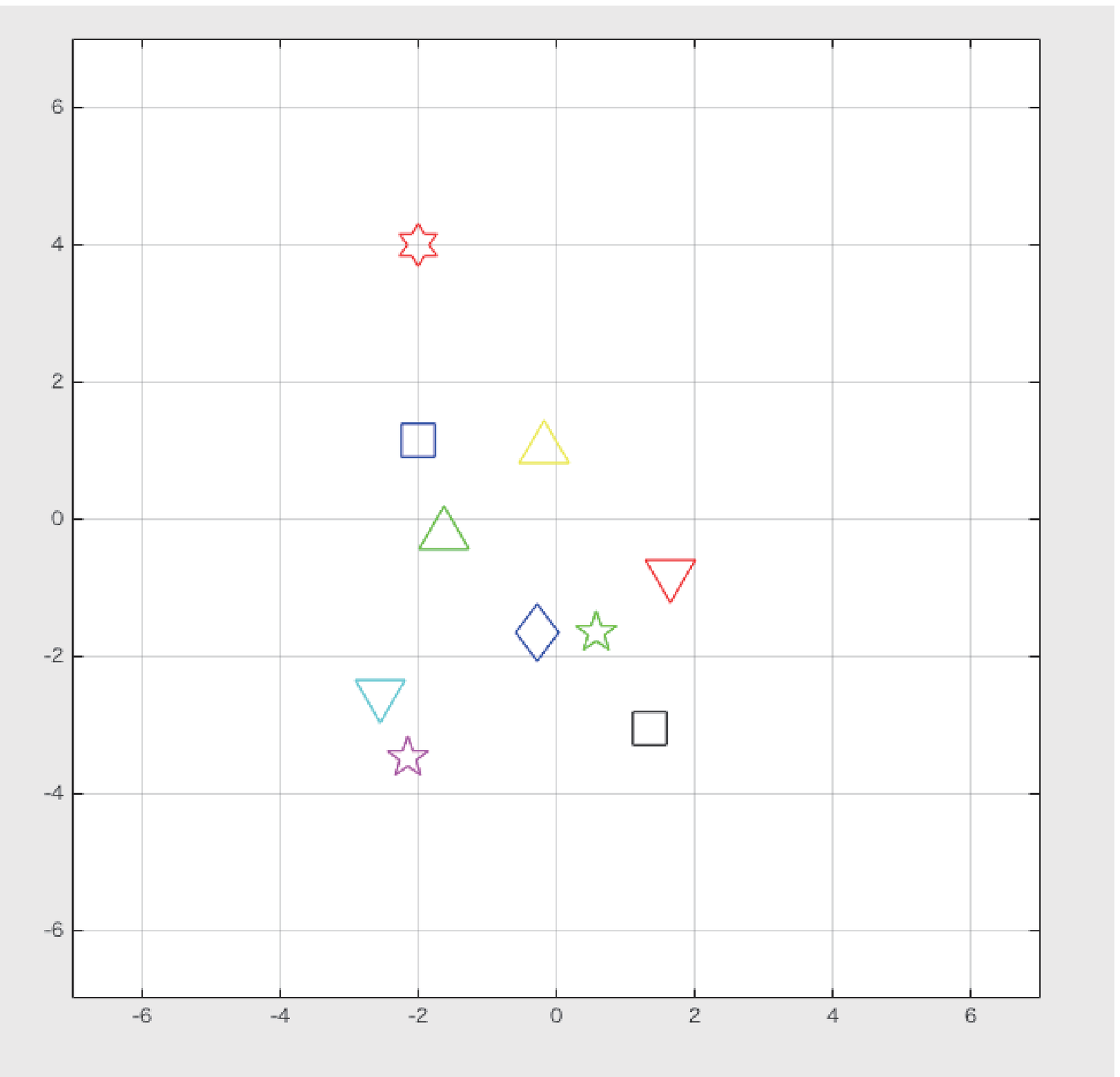}  
    \caption{$t=0.01$[$s$]}
  \end{subfigure}%
 \caption{Location of 10 agents: ($t=0$[$s$] and $t=0.01$[$s$])}
 \label{fig:t0.00and0.01}
\end{figure}

\begin{figure}[t]
  \centering
  \begin{subfigure}{.47\columnwidth}
    \centering
    \includegraphics[width=\linewidth]{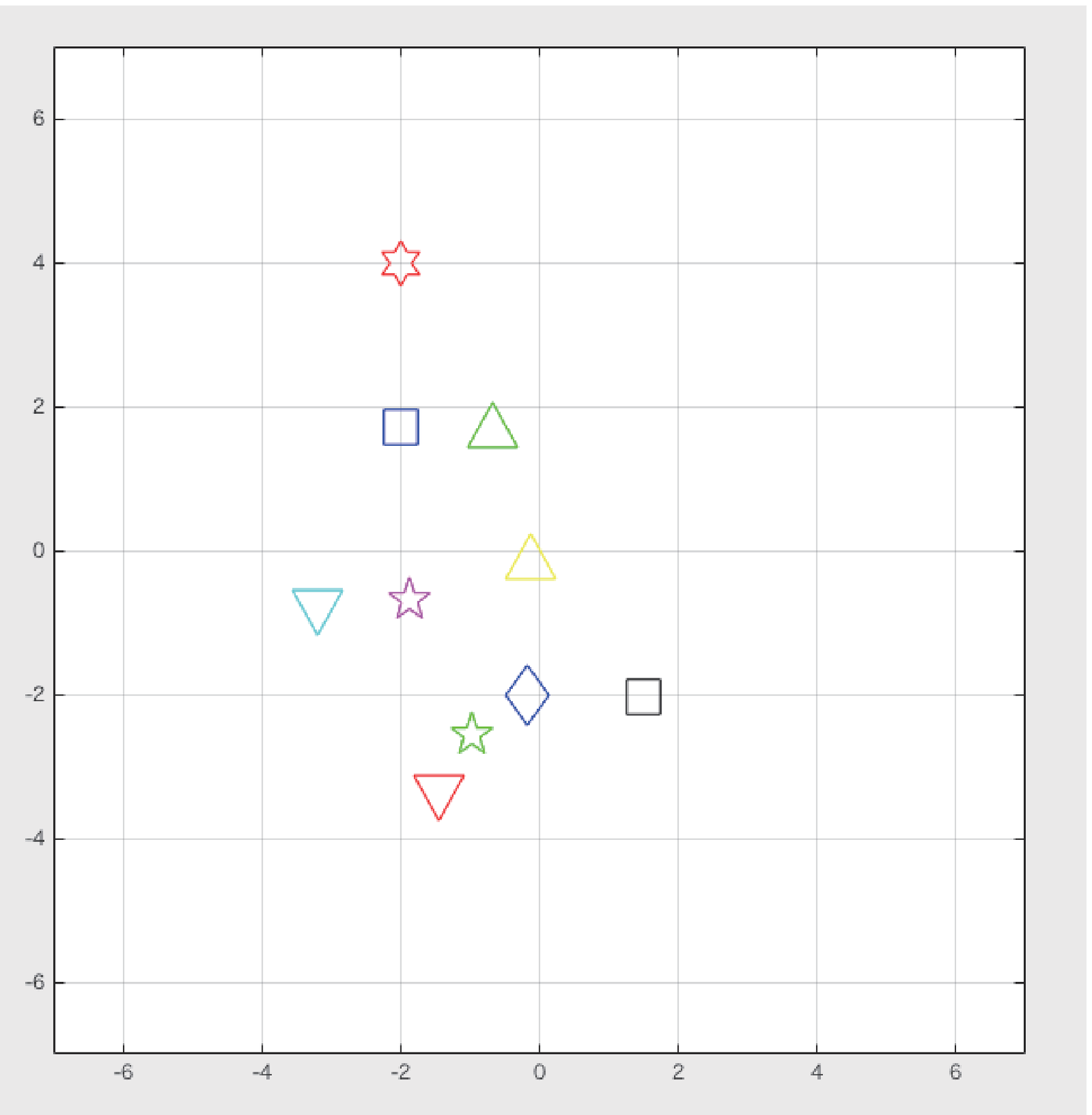}  
    \caption{$t=0.02$[$s$]}
  \end{subfigure}%
  \hfill
  \begin{subfigure}{.5\columnwidth}
    \centering
    \includegraphics[width=\linewidth]{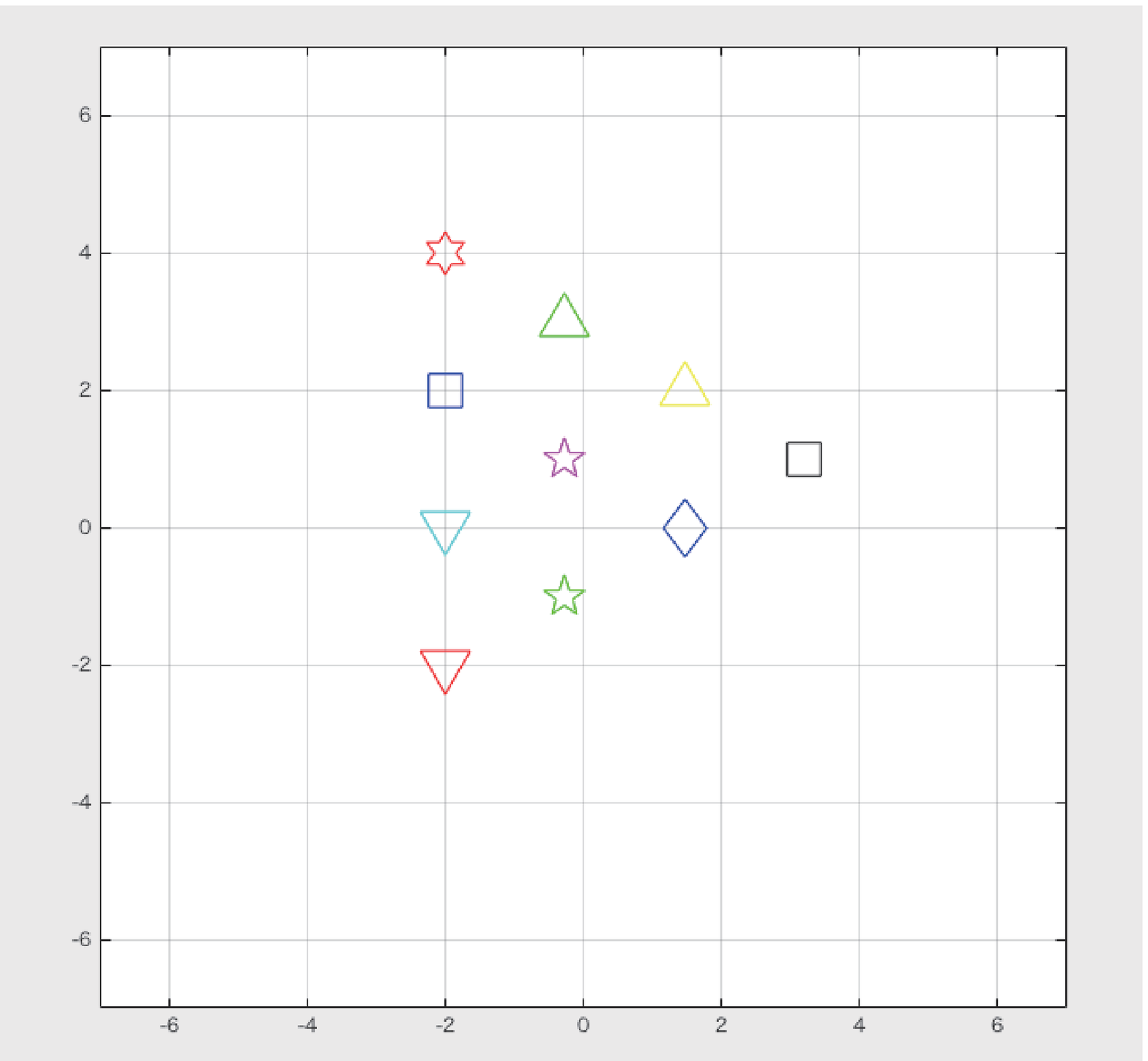}  
    \caption{$t=1$[$s$]}
  \end{subfigure}%
 \caption{Location of 10 agents: ($t=0.02$[$s$] and $t=1$[$s$])}
 \label{fig:t0.02and1}
\end{figure}

In this section, simulation results are shown to demonstrate the effectiveness of the proposed method. 
The multi-agent system and its desired formation are exactly the same as those in the previous section. The control gain used in the simulation is set as $K = 20$. 
Ten agents are located initially as shown in Fig.~\ref{fig:t0.00and0.01}(a). In the left column, agents $1 \sim 5$ are located from top to bottom. Agents $6 \sim 10$  are located similarly in the right column.
Figs.~Fig.~\ref{fig:t0.00and0.01}(b) and Fig.~\ref{fig:t0.02and1}(a) show snapshots of their locations at $t=0.01$ and $0.02$, respectively.
Fig.~Fig.~\ref{fig:t0.02and1}(b) shows the final formation of ten agents, which is the desired one.
It is also verified that the correct formation is achieved from various random initial locations.




\section{CONCLUSIONS}

This paper proposed a scalable formation shaping control based on the potential function  with distance and area constraints. It is able to achieve the desired distance between agents and the desired area constraints, i.e. there is no stable equilibrium involving flipping relative to the desired formation
shape. The proposed control strategy can be applicable for a triangulated formation with any number of agents (that can be constructed by Henneberg vertex extensions) in the case where all triangles are equilateral. A key analytical result is given, and some numerical results are shown to demonstrate the effectiveness of the proposed method. Current work is aimed at removing the restriction to equilateral triangles, which involves identifying the range of
acceptable gains involving the signed area in the relevant potential functions.

\begin{center}
{\bf Acknowledgement}
\end{center}
This work is supported by JSPS KAKENHI Grant number JP17H03281. The work of B. D. O. Anderson is supported by Data-61 CISRO, and by the Australian Research Council's Discovery Project DP-160104500. The authors thank Gangshan Jing for helpful discussions on this topic.


\begin{thebibliography}{99}
\bibitem{AYFH2008}
B.~D.~O. Anderson, C. Yu, B. Fidan, J. Hendrickx,  
Rigid graph control architectures for autonomous formations. IEEE Control Systems Magazine, Vol. 28, No. 6, pp. 48–63 ~ (2008)
\bibitem{ASSAS2017}
B. D. O. Anderson, Z. Sun, T. Sugie, S. Azuma and K. Sakurama,
Formation shape control with distance and area constraints, 
IFAC Journal of Systems and Control, Vol. 1, pp. 2-12~ (2017)
\bibitem{SAS2015}
K. Sakurama, S. Azuma, T. Sugie,
Distributed Controllers for Multi-Agent Coordination Via Gradient-Flow Approach, 
IEEE Trans. Autom. Control, Vol. 60, No. 6, pp.1471-1485~ (2015)
\bibitem{SAS2018}
K. Sakurama, S. Azuma, T. Sugie,
Multi-agent coordination to high-dimensional target subspaces, 
IEEE Trans. on Control of Network Systems , Vol. 5, No. 1, pp. 345-358 (2018.3) 
\bibitem{OPA2015}
K.K. Oh, M.C. Park, H.S. Ahn, 
A survey of multi-agent formation control. 
Automatica, Vol. 53 , pp.424-440~(2015)


\bibitem{olfati2003}
R. Olfati Saber, R. M. Murray, 
Flocking with obstacle avoidance: cooperation with limited information in mobile networks. 
Proc. of the 42nd IEEE Conference on Decision and Control,  pp. 2022-2028 (2003) 


\bibitem{zhao2018}
S. Zhao, D. V. Dimarogonas, Z. Sun, and D. Bauso. A general approach to coordination control of mobile agents with motion constraints. IEEE Trans. on Automatic Control vol. 63, no. 5 pp. 1509-1516 (2018) 

\bibitem{KBF2009}
L. Krick, M. E. Broucke, B. A. Francis,
Stabilization of infinitesimally rigid formations of multi-robot networks. 
International Journal of Control, Vol. 82, No. 3, pp. 423-429~(2009)

\bibitem{Sun2016}
Z. Sun, S. Mou, B. D. O. Anderson, and M. Cao. Exponential stability for formation control systems with generalized controllers: A unified approach. Systems \& Control Letters vol. 93, pp. 50-57~(2016)

\bibitem{O-S2006}
R. Olfati-Saber,
Flocking for multi-agent dynamic systems: algorithm and theory,
IEEE Trans. Autom. Control, Vol. 51, No. 3, pp. 401-420~(2006) 
\bibitem{CBB2017}
X. Chen, M.-A. Belabbas, T. Basar,
Global stabilization of triangulated formations,
SIAM Journal of Control Optim., Vol. 55, No. 1, pp. 172-199~(2017) 
\bibitem{AK2006}
P.-A. Absil and K. Kurdyka,
On the stable equilibrium points of gradient systems,
Systems {\&} Control Letters, Vol. 55, No. 7, pp.573-577~(2006) 
\bibitem{Terrell2009}
W. J. Terrell, Stability and stabilization: An introduction,
Princeton University Press. (2009)
\end{thebibliography}
\end{document}